\documentclass[]{aa}
\usepackage{array}
\usepackage{natbib}

\usepackage[dvipsnames]{xcolor}
\usepackage{graphicx}
\usepackage{epstopdf}
\usepackage{mwe}
\usepackage{caption}
\usepackage{subcaption}
\usepackage[varg]{txfonts}
\usepackage[bookmarks = true,colorlinks = true,linkcolor = red,urlcolor = blue,citecolor = blue,pdftex,bookmarks = true,linktocpage = true,hyperindex = true,breaklinks]{hyperref}
\usepackage{longtable} %for huge tables
\usepackage{fancyvrb}
\usepackage{float} % the sideways table package seems to need this for me to able to export to arxiv  

\usepackage{multirow}
\usepackage{lscape}
\usepackage{ulem}
\usepackage{rotating}

\usepackage{colortbl}
\definecolor{Grey}{rgb}{0.5,0.5,0.5} 
\definecolor{MyRed}{rgb}{0.9,0.0,0.0} 
\definecolor{MyPink}{rgb}{0.8,0.3,0.5} 
\definecolor{MyMediumBlue}{rgb}{0.7,0.72,1.0} 
\definecolor{MyGreen}{rgb}{0.0,0.9,0.5} 
\definecolor{SWGreen}{rgb}{0.0,0.5,0.0} 
\definecolor{SWRed}{rgb}{0.9,0.0,0.0} 
\definecolor{AMpurple}{rgb}{0.9,0,1.0} 
\definecolor{AMbrown}{rgb}{0.8,0.3,0.3}

\begin{document}

   \title{The Solar ALMA Science Archive (SALSA)}
\subtitle{First release, SALAT and FITS header standard.}
  % \subtitle{The first science-ready mm-imaging resource for solar physicists and respective tools}

  \author{Vasco M. J. Henriques\inst{1}$^,$\inst{2}
        \and 
    Shahin Jafarzadeh  \inst{1,2}
            \and 
    Juan Camilo Guevara Gómez   \inst{1,2}
            \and 
    Henrik Eklund  \inst{1,2}
  \and 
      Sven Wedemeyer \inst{1,2}
    \and
  Miko\l{}aj Szydlarski \inst{1,2}
    \and
    Stein Vidar H. Haugan\inst{1,2}
        \and
   Atul Mohan\inst{1,2}
          }

  \institute{Rosseland Centre for Solar Physics, University of Oslo, P.O. Box 1029 Blindern, NO-0315 Oslo, Norway
          \and
    Institute of Theoretical Astrophysics, University of Oslo, P.O. Box 1029 Blindern, NO-0315 Oslo, Norway\\
             \email{vh@astro.uio.no}
             }

% \abstract{}{}{}{}{} 
% 5 {} token are mandatory
 \date{XXXX; accepted XXXX}
 
  \abstract
   {In December 2016, the Atacama Large Millimeter/submillimeter Array (ALMA) carried out the first regular observations of the Sun. These early observations and the reduction of the respective data posed a challenge due to the novelty and complexity of observing the Sun with ALMA. 
   The difficulties with producing science-ready time-resolved imaging products in a format familiar and usable by solar physicists based on the measurement sets delivered by ALMA had so far limited the availability of such data. With the development of the Solar ALMA Pipeline (SoAP), it has now become possible to routinely reduce such data sets. 
   As a result, a growing number of science-ready solar ALMA datasets is now offered in the form of Solar ALMA Science Archive (SALSA).  
   So far, SALSA contains primarily time series of single-pointing interferometric images at cadences of one or two seconds. The data arrays are provided in FITS format.  
   We also present the first version of a standardised header format that accommodates future expansions and fits within the scope of other standards including the ALMA Science Archive itself and SOLARNET. The headers also include information designed to aid the reproduction of the imaging products from the raw data. Links to co-observations, if available, with a focus on those of the Interface Region Imaging Spectrograph (IRIS), are also provided. SALSA is accompanied by the Solar ALMA Library of Auxiliary Tools (SALAT) that contains IDL and Python routines for convenient loading and quick-look analysis of SALSA data.}

 \authorrunning{}

\maketitle
%

%-------------------------------------------------------------------

\section{Introduction}
\label{Section:Introduction}
%(captured in the Einstein coefficients)
%\swrm{A prime physics laboratory, the Sun provides us with a continuous real-time display of complex physical interplay across multiple scales where gas pressure and magnetic pressure meet and trade places as the dominant driver, in space an time, touched by reconnection and non-local radiation effects. Arguably the richest region of the solar atmosphere is the chromosphere. Part of this rich physics are the energy level populations of the atoms and ions that offer us precious empirical window that spans the UV to the IR. These are affected by a multitude of effects which include local heating via collisions, and incoming and outgoing radiation that can be highly non-local. Fundamental in the formation of the observed line-profiles, these effects are both diagnostic and unknown variables. As variables they can be completely unconstrained and time dependent. For example when wave/shock dynamics cause local density changes lasting seconds,  changing how much of the energy levels are due to collisions and thus due to the local temperature conditions, or when a nearby event changes the radiation field significantly causing departures form local conditions. Known as non-local thermal equilibrium (non-LTE) effects, these physical interplays cause a fundamental limitation on our capacity to constrain the physical conditions of the solar chromosphere.} 

The millimetre wavelength range offers a unique window into the chromosphere of the Sun. In contrast to other chromospheric diagnostics such as spectral lines that are typically affected by deviations from local thermodynamic equilibrium (LTE) conditions, the assumption of LTE is believed to be valid for the radiation continuum at millimetre wavelengths as it is formed by thermal bremsstrahlung (free-free emission) \citep[see, e.g.,][and references therein]{2021MNRAS.500.1964V,2016SSRv..200....1W,2006A&A...456..697W,1985ARA&A..23..169D}. The formation process of the radiation allows to use the Rayleigh-Jeans approximation, providing a linear relation between the measured flux density and the brightness temperature, which is under ideal conditions closely connected to the local plasma temperature in the continuum-forming atmospheric layer. A particular complication with observing the chromosphere is that it evolves on dynamical time scales of seconds, and exhibits ubiquitous fine-structure on sub-arcsec spatial scales. 
Resolving both simultaneously is technically challenging, especially in view of the relatively long wavelengths and the resulting need of correspondingly very large telescope apertures. 
The technical requirements, which are essential for exploiting the full potential of the millimetre window for the solar chromosphere, are currently only met by the Atacama Large Millimeter/sub-millimeter Array (ALMA; \citealt{2009IEEEP..97.1463W}). 
Regular solar observations are performed by ALMA since 2016. For this purpose, the 12-m Array with up to fifty antennas, each with a diameter of 12\,m, is combined with the Atacama Compact Array (ACA) with up to twelve 7-m antennas. The combined interferometric array is sensitive to brightness temperature variations over spatial scales and orientations according to the array configuration relative to the position of the Sun on the sky. Absolute brightness temperatures are derived by combining the images reconstructed from the interferometric data with additional single-dish scans of the solar disc with up to four Total Power  
(TP) antennas. 
Please refer to \citet{2017SoPh..292...87S} for details on the interferometric part and to \citet{2017SoPh..292...88W} for more information on the single-dish TP component   
\citep[see also, e.g.,][and references therein]{2002AN....323..271B,2008Ap&SS.313..197L,2011SoPh..268..165K,2018Msngr.171...25B}.

Despite the technical challenges, an increasing number of studies based on ALMA observations of the Sun is being published 
\citep[see,  e.g.][ and more]{2017ApJ...845L..19B,
2017ApJ...841L...5S,
2018ApJ...863...96Y,
2018A&A...613A..17B,
2019A&A...622A.150J,
2019ApJ...875..163R,
2019ApJ...871...45S,
2020A&A...635A..71W,
2020A&A...643A..41D, 
2020A&A...644A.152E,
2020A&A...638A..62N, 
2020A&A...634A..86P, 
2021A&A...651A...6B,
2021ApJ...906...82C,
2021RSPTA.37900184G,
2021RSPTA.37900174J,
2021A&A...652A..92N}.
While scientific production using ALMA solar data is clearly picking up, interferometric imaging into stabilised time-consistent data series in absolute temperature units remained difficult. 
Due to the significant differences between solar observations and standard ALMA observations of other astronomical targets, ALMA so far delivers calibrated measurement sets to the observers, or more precisely, the data together with a prepared calibration script that is to be executed by the recipient. 
The further processing, which includes the construction of time series of the images, requires significant experience, time and resources, which has hampered access for solar physicists who are yet unfamiliar with millimetre data. 
This difficult situation initiated the development of the Solar ALMA Pipeline \citep[SoAP,][]{soap} with the aim to provide easy access to science-ready solar ALMA data to the scientific community. 
Routine data processing with SoAP has so far resulted in multiple science-ready imaging time-series. 
The original data sets were retrieved from ESO's ALMA Science Archive 
\citep[ASA\footnote{\url{https://almascience.eso.org/aq/}},][]{2017Msngr.167....2S} 
from where they can be freely downloaded after the end of the proprietary period. 
%\todo{Extensive use of the highly-searchable ESO ASA (cite) was made.} 
For the publicly available sets, the calibration scripts were executed and SoAP was applied. All resulting processed data sets that were of sufficient quality are now 
%With re-calibration and reduction performed from scratch on all publicly available sets, a selection of science usable sets are 
made available online in Solar ALMA Science Archive (SALSA), which is complementary to ASA in the sense that it contains science-ready data. 

Here we present the first version of SALSA, consisting of 26 such sets, the accompanying Solar ALMA Library of Auxiliary Tools (SALAT), and the first version of a header standard for the included FITS data files. 
The observational data sets included in the current version of SALSA are briefly described in Sect.~\ref{sec:observation}. A brief summary of the data processing with SoAP and the resulting data products are provided in Sects.~\ref{sec:process} and \ref{sec:dataproducts}, respectively. 
SALSA and SALAT together with guidelines for usage, documentation and acknowledgements are addressed in Sect.~\ref{sec:salsa-salat}, followed by a summary and outlook in Sect.~\ref{sec:outlook}.

%The primary goal of the archive is to provide a quick jumpstart to solar science using the unique properties of mm interferometric observations. Together with the archive we provide a set of tools to assist with the manipulation and analysis thereof.   

\begin{figure}[t!]
    \centering
    \includegraphics[width=\columnwidth]{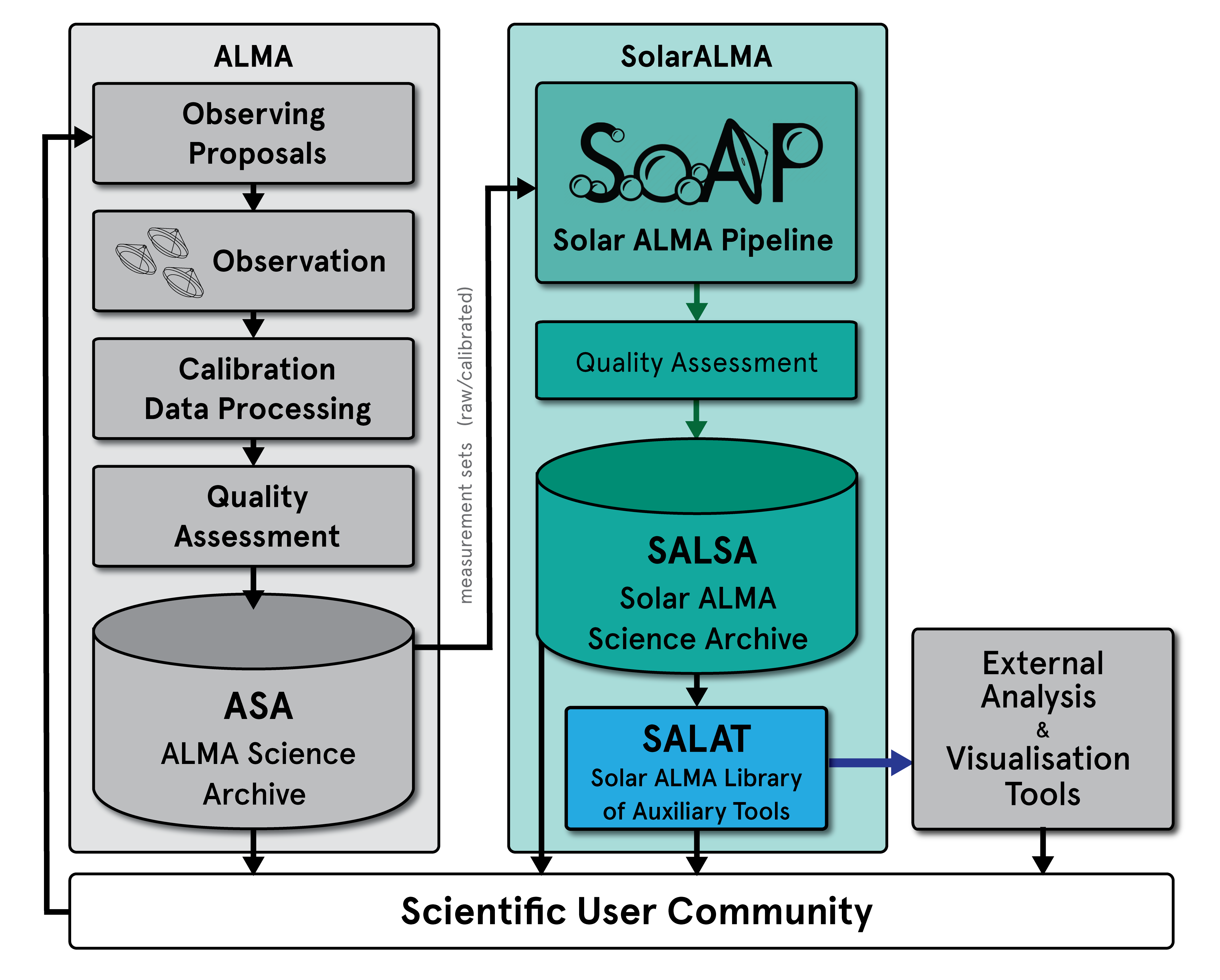}
    \caption{While scientists can always download solar measurements sets directly from the ALMA Science Archive, the SolarALMA project created an alternative, which makes access to public, science-ready data easier. The Solar ALMA Science Archive (SALSA) is accompanied with a tool library (SALAT) that also allows for exporting SALSA data into a format that can be other analysis and visualisation tools. Please see the main text for more details.}
    \label{fig:my_label}
\end{figure}

\section{Observations}
\label{sec:observation}

 %This suggested absolute level for most sets comes from the temperature measure by a single antenna at the region of the observed field-of-view. 
%The files are compliant with the Flexible Image Transport System NASA standard version 3.0, 4.0, with the evolving ALMA header format standard, and the SOLARNET Metadata Recommendations document version 1.4. %should change to 1.5

Solar observations are performed using an heterogeneous array through the combination of the 12-m Array and the ACA with additional, typically simultaneous single-dish full-disk scans of the Sun with several TP antennas. The data provided in the first version of SALSA presented here was obtained in the period from December 2016 to 2018. The observations were performed in Band~3, covering the frequency range between 92 and 108\,GHz, and Band~6 with frequencies between 229 and 249\,GHz. These frequency ranges correspond roughly to an average wavelength of 3~mm for Band~3 and 1.3~mm for Band~6, respectively. The radiation in these bands is dominated by (thermal) free-free continuum emission. 

\begin{figure}[t!]
    \centering
    \includegraphics[width=\columnwidth]{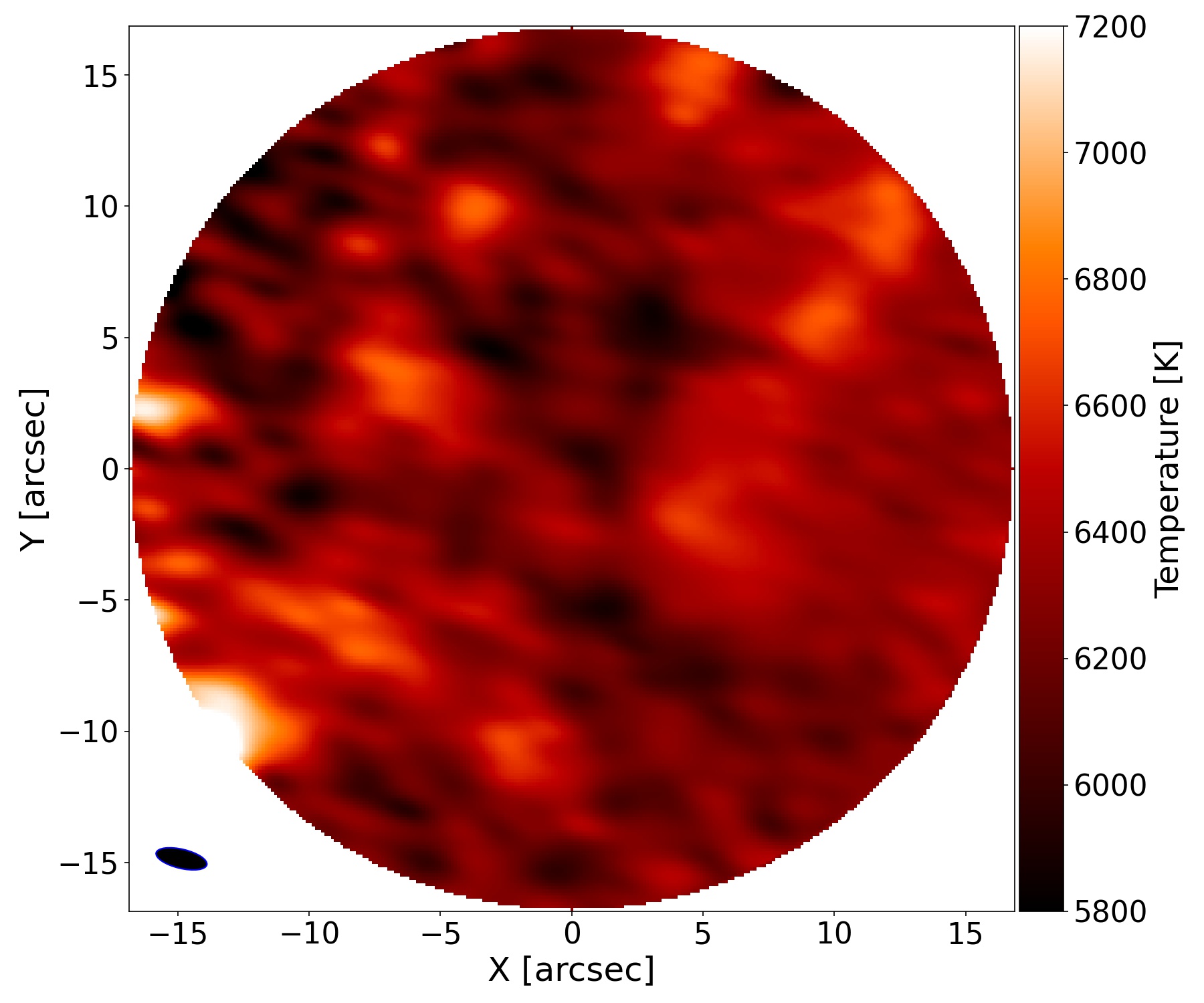}
    \caption{A generic frame from a disk centre set (D22), showing quiet Sun and network features as observed in Band~6. The elliptic cross section of the beam shape is plotted in the bottom left of the frame. This plot has been made with SALAT (see Section~\ref{sec:salat}).}
    \label{fig:generic_frame}
\end{figure}

The datasets in SALSA are labelled as Dnn (in this first release D01 to D28; see Table~\ref{tab:sets}). We refer to these labels when discussing slight differences in the sets below. 
All released sets are constructed from single-pointing observations with 12~m and ACA and contain  time-sequences of absolute brightness temperature maps (in units of Kelvin) at a cadence of 1 or 2~seconds 
spanning from a few minutes to an hour of Solar evolution. The 26 data sets released here are part of the following 11 successful observational projects: 
%These projects are, as labelled following the ADS/JAO ALMA project identification (Project ID) numbers: 
2016.1.00030.S, 2016.1.00050.S, 2016.1.00202.S, 2016.1.00423.S, 2016.1.00572.S, 2016.1.01129.S, 2016.1.01532.S, 2017.1.00653.S, 2017.1.01672.S, 2018.1.01879.S, and 2018.1.01763.S.
The individual observations target quiet Sun and active regions for a variety of heliocentric angles on the solar disk.   
An illustrative example of a resulting snapshot in Band~6, close to the disk centre (from D22), is shown in Fig.~\ref{fig:generic_frame}. 
%as is the lower chromospheric "fluctusphere". 
All time-series display notable evolution for all visible features, including fibrils in some datasets. 
The included limb dataset features a spicule or small protuberance (set labelled as D11). 
The effective angular resolution of the datasets, as set by the synthesised beam size, which corresponds to the primary lobe of a point-spread function, depends mainly on the wavelength, configuration of the interferometric array and position of the target on the sky. Consequently, the effective angular resolution varies significantly for the individual datasets and also from frame to frame within the individual time series due to the motion of the Sun on the sky and various technical reasons (e.g., flagging of individual antennas). 
%and provided as part of the data. 
% The datasets presented here are observed in different array configurations. The effective angular resolution of the datasets therefore varies significantly per wavelength. The pixel size is optimized to oversample the resolution and ranges from 0.14~arcsec to 0.35~arcsec. A close proxy of the actual resolution: the beam size, for which an analogous in the optical would be the point-spread function, is variable per frame and provided as part of the data. 

%\begin{figure}[t!]
%    \centering
%    \includegraphics[width=\columnwidth,trim=0 0 50 0,clip=true]{figs/context2.eps}
%    \caption{A generic frame from a disk centre set (D22), showing quiet Sun and network features as observed in Band~6. }
%    \label{fig:my_label}
%\end{figure}

\begin{figure*}[!thp]
\centering
    \includegraphics[width=.80\textwidth, trim = 0 0 0 0, clip]{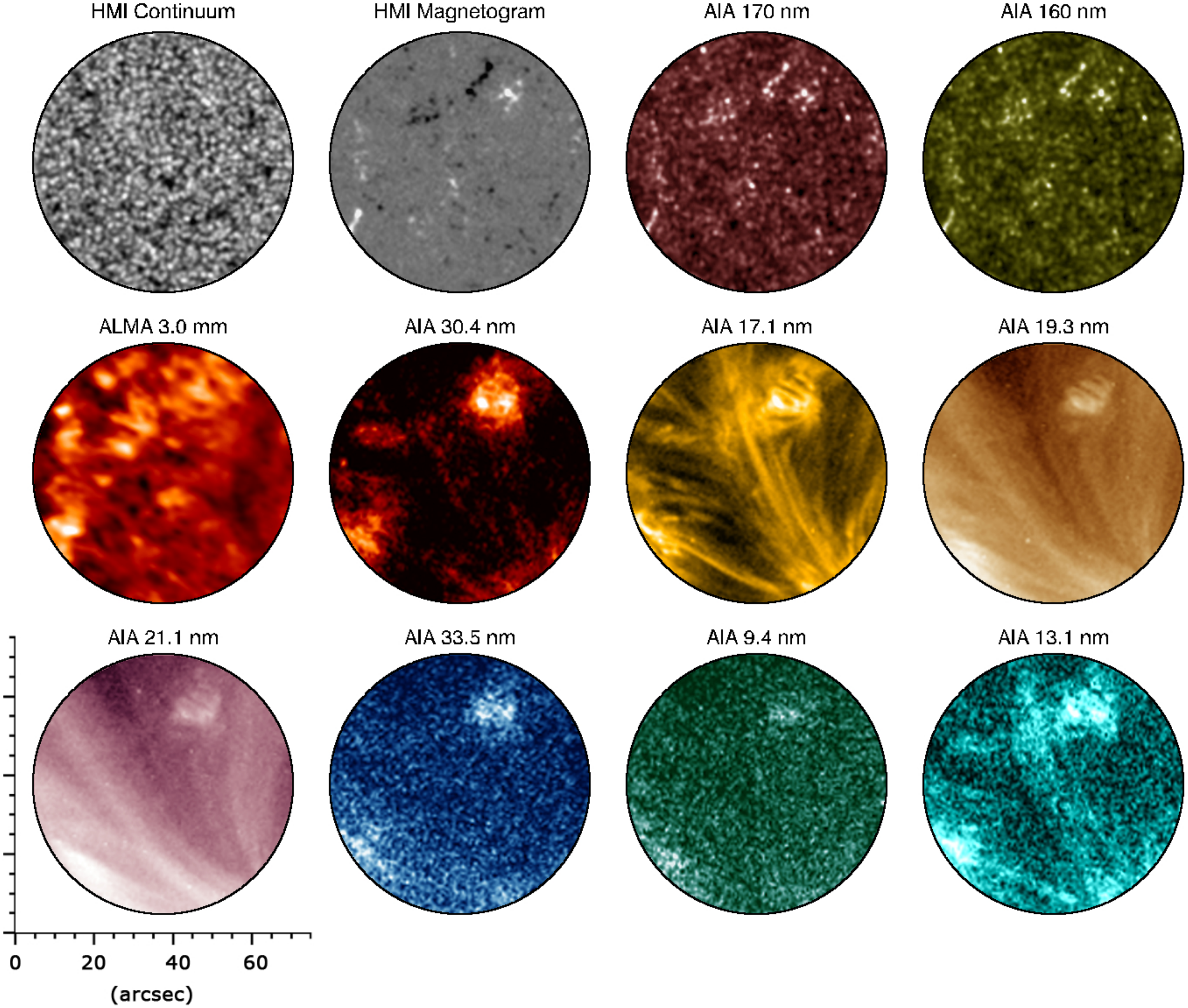}
  \caption{An ALMA Band 3 image from 2016-12-22 (project ID 2016.1.00423.S), along with co-aligned SDO images. Time series of images as a movie is available online.}
  \label{fig:b3_20161222}
\end{figure*}

The ALMA datasets can be aligned (spatially and/or temporally) with observations from other ground-based and/or space-borne telescopes (at other wavelengths) to further inspect, e.g., the evolution of features of interest at multiple atmospheric heights and for general context. In particular, the Solar Dynamics Observatory (SDO; \citealt{2012SoPh..275....3P}) with continuous observations of the entire solar disc at several wavelength bands (sampling the solar photosphere, low chromosphere, transition region, and corona) can provide complementary information to the ALMA observations (these complement each other since SDO does not sample the chromospheric heights captured by ALMA). Figure~\ref{fig:b3_20161222} shows an example of such alignments between an ALMA image and its co-tempo-spatial images from the Helioseismic and Magnetic Imager (HMI; \citep{2012SoPh..275..229S} and the Atmospheric Imaging Assembly (AIA; \citealt{2012SoPh..275...17L}) (onboard SDO) for the first SALSA dataset (i.e., D1), taken on 2016-12-22 in Band~3. All SALSA sets are provided with helioprojective coordinates which should facilitate such alignments. However, the given coordinates and position angles (which allow the rotation of the field of view with respect to the solar north-south direction) may be used only as a first approximation, since offsets can be expected (see Sect.~\ref{sec:process} for details). It was found that a combination of the SDO/AIA 170~nm and 30.4~nm channels (with different brightness weights) resulted in a similar scene to that sampled by ALMA (in both Band 3 and 6). For Band 3, adding the SDO/AIA 17.1~nm channel was helpful in some cases. The combined image facilitates cross correlations between similar solar features with ALMA observations. The field of view illustrated in Figure~\ref{fig:b3_20161222} samples a magnetically quiescent area, with small network patches (of opposite polarities) as seen in, e.g., the HMI magnetogram and the AIA 170 and 160~nm channels. Excess brightness temperature in the ALMA image (at 3.0 mm; Band 3) over the network patches as well as intensity enhancements in hotter AIA channels, particularly loops connecting the opposite polarities, are evident. Larger loops, entering the field of view are rooted in strong magnetic concentrations outside the target area. See, e.g., \citet{2020A&A...635A..71W} and \citet{2021RSPTA.37900174J}, for detailed analyses of this  dataset.

\begin{figure*}[!thp]
\centering
    \includegraphics[width=.99\textwidth, trim = 0 0 0 0, clip]{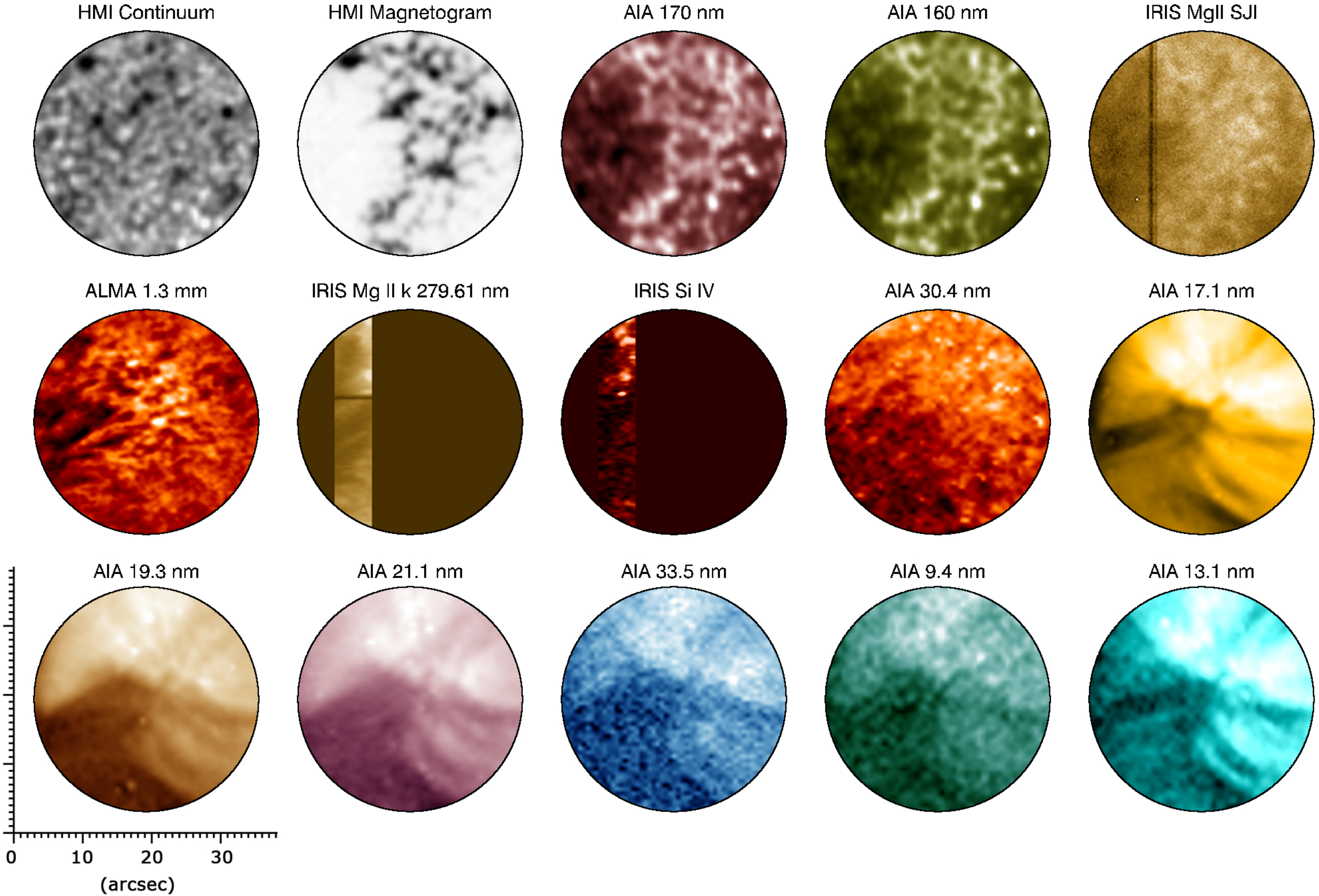}
  \caption{Same as Figure~\ref{fig:b3_20161222}, but for ALMA Band 6 taken on 2017-04-22 (project ID 2016.1.00050.S). In addition to the SDO channels, one slit-jaw image (SJI) and two raster images from IRIS are also displayed. Time series of images as a movie is available online.}
  \label{fig:b6_20170422}
\end{figure*}

Another example of co-observations with other instruments is presented in Fig.~\ref{fig:b6_20170422} where the co-spatial images from the SDO channels and selected images from the Interface Region Imaging Spectrograph (IRIS; \citealt{2014SoPh..289.2733D}) explorer are shown along with the ALMA Band~6 image (from 2017-04-22; D08) at the beginning of the observations. This dataset samples a plage/enhanced-network region, with excess brightenings above the magnetic concentrations in the ALMA image, while the chromospheric (dark) fibrillar structures (over the internetwork area) are also observed in both ALMA and IRIS Mg~{\sc ii}~k at 279.61~nm. The latter is a raster image where the slit has scanned a relatively small region compared to that of ALMA. An IRIS slit-jaw image (SJI) at around the Mg~{\sc ii} line is also shown in~Fig.~\ref{fig:b6_20170422}. 

The time-series of images as shown in  Figs.~\ref{fig:b3_20161222} and \ref{fig:b6_20170422} are available online as a movie (please see the online material connected to this article). We note that since the ALMA and SDO images were taken with different cadence, the ALMA observations with 2~sec cadence was considered as the reference. Hence, the SDO (and IRIS) images are repeated in time to fill in the gaps, in a way that the time difference of the images are the shortest at any given time. These will result in co-alignments of the images both temporally and spatially. It is worth noting that the SDO images in these two examples were resampled to the pixel size of the ALMA images prior to the alignment, but they were not convolved with the synthesised beam of ALMA (i.e. the corresponding PSF). For a one-to-one comparison, such convolutions may be necessary. Necessary routines for extracting a desired ALMA PSF and performing convolutions, along with other useful simple analyses, are provided through the Solar ALMA Library of Auxiliary Tools (SALAT; see Sect.~\ref{sec:salat}).

\section{Data processing}
\label{sec:process}
%As mentioned above, the primary goal behind SoAP and SALSA is to simplify processing calibrated ALMA measurement sets into science-ready FITS imaging data. 

This first release of SALSA is the product of routine application of the Solar ALMA Pipeline (SoAP) to the publicly released measurement sets available on the ALMA Science Archive \citep[ASA,][]{2017Msngr.167....2S}. As mentioned above, the primary goal behind SoAP is to simplify processing calibrated ALMA measurement sets into science-ready FITS imaging data. For all data sets the following steps of SoAP were executed:
\begin{enumerate}
    \item Imaging including CLEAN wrapper, self-calibration (for all but one data set) and primary beam correction
    \item Brightness temperature conversion (from Jy/beam to K)
    \item Bad frame discarding in time series
    \item Frame-to-frame spatial alignment for ALMA time sequences
    \item Combination of interferometric and Total Power data
\end{enumerate}

The imaging part of SoAP uses the CLEAN algorithm \citep{1974A&AS...15..417H} for the deconvolution of images in the form devised by  \citet{2011A&A...532A..71R} and as implemented in CASA \citep[Common Astronomy Software Applications;][]{casa}. The conversion from flux density to brightness temperature was done using the Rayleigh-Jeans approximation of the Planck function and the time-dependent solid angle (i.e. the area covered on the sky) of ALMA's synthesised (elliptical cross-section) beam, following the standard procedure\footnote{\url{https://science.nrao.edu/facilities/vla/proposing/TBconv}}.% Typical conversion factors range from 5 to 40 \todo{K Jy$^{-1}$ arcsec$^{2}$ ... }. %assuming that the synthesised beam response should have a two-dimensional Gaussian profile with an elliptical cross section. Typical conversion factors range from 5 to 40. 

For all sets except D01, self-calibration for phase \citep{cornwell99} was performed. For D01 the self-calibration method severely under-performed due to technical problems with pointing for those early observations. Spatial alignment (single shift value per frame, i.e. rigid) was performed for all sets except the limb set to effectively correct any residual jitter and wobbling of the time-series. This follows an heavily adapted version of CRISPRED's polish tseries core routines \citep{2015A&A...573A..40D}. Visually, the alignment procedure is generally successful in keeping features aligned and thus maximises science potential for time evolution studies. 

The combining of total power and interferometric data was performed in different ways as the pipeline development progressed. The primary goal of this step is to obtain \textit{absolute} brightness temperatures since the interferometer only provides differences (as the zero frequency of the Fourier space would require a zero distance between two antennas which cannot be obtained). For D01 and D08 this combination follows a process known as ``feathering'' and described in \citep{2002ASPC..278..375S}. 
This procedure consisted of rescaling a matching TP map to the interferometric pixel scale and adding those values pixel-wise (spatially matched) to the entire cube. This technique allows for compensation of the incomplete uv Fourier space sampling by the interferometer by adding the low frequency spatial variation at the same time as the absolute temperature. This technique has been favoured also in \citep{2021ApJ...906...83C} as it is desirable to recover the low frequency spatial scales. However, the interferometric maps are available at a cadence of 1-2\,s while only one single or a few TP maps with many minutes in-between are usually available. Consequently, the lower spatial frequency scales, set by the TP, will be steady while the high spatial frequency features as set by the interferometer evolve. This would impact the mapping of features evolving on the affected scales with the effect becoming more severe for larger differences between the time at which the interferometric data was taken and the time at which the employed TP data were recorded. %, becoming more of an issue for the longer time series. 
%This is also an issue for when the TP is not acquired close enough in time to the centre of the time series. 
For this reason, save the few exceptions highlighted both here and in the metadata (see Sect.~\ref{sec:dataformat}) of the data files, the combining was done by computing an average over 3x3 TP pixels at the location of the interferometric observations and merely adding it to the interferometric data. This simpler technique avoids the introduction of false low spatial frequency structures but reduces the low spatial frequency information. Depending on the science case, this is either an advantage or a disadvantage and a user of interferometric data, including the one put forward in this database, should be aware of the different choices made for each set. Aside from this paper, the information on this step, can be found in extension 'PRPARA' (extension~2) of each FITS data file. %Sets labelled D02 to D10 were processed with an earlier version of the pipeline, for these, the offsets were set to a quiet-Sun value, also listed in the headers, of either 7300~K for Band~3 or 5900~K for Band~6,
For the few data sets for which no TP single-dish observations were available, the offsets were set to the default (quiet-Sun) value, also listed in the headers, of either 7300~K for Band~3 or 5900~K for Band~6 as suggested by \citet{2017SoPh..292...88W}. The same was done for the limb set (D11) as the TP offset setting close to the limb is highly uncertain. The latter is due to an overshooting pattern present when scanning the disk with a single dish and to small pointing differences leading to large changes in absolute observed temperature. Thus the present limb set and any future releases of limb sets should only be used to study variations in temperature until the issue has been fixed. 
Each TP, as available in ASA, is itself calibrated by assuming that the average brightness temperature in a region around disk centre has a value of 7300~K for Band~3 or 5900~K for Band~6, again following \citet{2017SoPh..292...88W}. Note that this is not the same as hard setting the total power for an interferometric time series calibrated with that TP since, depending on activity levels and heliocentric angle, the offset extracted from the TP will be different. 

Multiple scans, i.e. continuous observing periods, have been combined whenever possible provided the target is the same and the time difference between scans is small. Because small time gaps are present between scans, the user should use the time stamps per frame when doing any time-dependent analysis to avoid any such gaps or otherwise missing frames (e.g. filtered-out bad frames) within the series. 

A final step to further reduce high frequency noise consisted of smoothing in time. The smoothing function and size are listed in the FITS metadata. For this release most sets received a final smoothing with a boxcar of 5 frames (so 5 to 10 second smoothing depending on the set cadence). This smoothing was found to be important to be able to release all time series at the highest cadence possible but ongoing refinements may make such filtering obsolete for most sets in future releases. 

Coordinates of the observations are computed using the RA and DEC listed in each measurement set released by the ASA and the ``reference time'' (REFTIME), which is usually the time of the first solar pointing (first same-named source for the first science scan in the case of multiple solar targets). These measurement set coordinates are converted to helioprojective Cartesian (X,Y) coordinates in arcsec by using the solar disk centre RA and DEC at the reference time. SoAP includes a routine for such conversions but, for some data sets in this first release, the web interface of the ALMA Solar Ephemeris Generator \citep{2019arXiv190408263S}\footnote{\url{https://celestialscenes.com/alma/convert/}}  was used. 
%from the ALMA Czech node was also used for this task (cite or footnote). 
The helioprojective coordinates, the reference time, and other useful coordinate information such as the ``P'' angle (i.e., position angle of the solar north pole) are listed in the headers of each data file under the World Coordinate System A (see Appendix~\ref{sec:datahead}). Due to  pointing issues in the early cycles, a target dependent error of up to 20~arcsec is expected to be present (private communication). The pointing accuracy was improved for sets where co-simultaneous observations (e.g. with SDO)  allowed for a refinement of the actually observed coordinates. The listed coordinates are the corrected values whenever possible, but in principle, they should be taken with caution due to the possibility of persisting multi-arcsec offsets from actual coordinates. %This was the case for D03, D04 and D05. %this consisted of visually adjusted using comparison with multiple SDO channels including HMI continuum and AIA 1600, 304 bandpasses.

\section{Data products}
\label{sec:dataproducts}
SALSA is hosted\footnote{\url{http://sdc.uio.no/salsa/}} by the Science Data Centre at the University of Oslo, which also, among other things, provides Hinode data. In this section, we describe the general format and content of the data products and the selection of data sets provided in the first release of SALSA.

\subsection{Data format and header standard}
\label{sec:dataformat}

The ALMA data on SALSA is provided as Flexible Image Transport System (FITS) files containing the brightness temperatures as five-dimensional data arrays (in units of Kelvin) and two extensions. The first two array dimensions are the usual spatial dimensions, the third is wavelength, fourth Stokes, and fifth time. This format was chosen in preparation of future data sets that will make use of all five dimensions, even though the data sets in this first release consist of continuum time series only (one wavelength, one Stokes component). 
Extension 'TIMEVAR' contains a binary table with time variable information. The table is self described in the headers of the extension and will be extended in future releases. At the time of this release, it consists of the beam properties and time tags for each frame present in the cube. The first element of the table is the major axis of ellipse used to describe the beam, the second the minor axis, the third is the beam position angle and the fourth the time tag. Following \cite{2015A&A...574A..36R} the time tags are presented in seconds counting from DATEREF (not to be confused with REFTIME), here set to the midnight at the start of the date of the observations (i.e. 00:00:00.000). These beam properties can be used to compute a proxy for angular resolution by taking the mean of the minor and major axis and such information is provided in the headers under the ALMA keyword SPATRES. For this database SPATRES varies from 0.77 arcsec to 4.11 arcsec. The time dependent beam properties from extension TIMEVAR should be considered for wave studies where very small scales are analysed. 

The second extension, PRPARA, contains an ASCII table with parameter names, values, and descriptions for different processing steps. The processing steps are named in the header of the extension and also in the primary header of the file using SOLARNET's "detailed description of all processing steps" formalism (Sect.~8.2 thereof) which can be identified by the PRXXXn keywords. These include information on  the used versions of CASA, SoAP, and other relevant details concerning data processing. Together with the public release of SoAP, which is planned for the near future, this allows advanced users to not only reproduce the data but also potentially optimise the data reduction towards wanted properties, e.g. prioritising brightness temperature over accurate reproduction of spatial structure. Further details on the headers and extensions is provided in the Appendix~(\ref{sec:datahead} and \ref{sec:extension_data}).
The files are compliant with the Flexible Image Transport System NASA standard version 3.0, 4.0, with the evolving ALMA header format standard (Felix Stoehr, private communication), and the SOLARNET Metadata Recommendations document version 1.4 \citep{2020arXiv201112139H}. 

\subsection{Data sets in the first SALSA release}
\label{sec:data}

All data sets that are included in the first release of SALSA  are listed in Table~\ref{tab:sets}. They all contain time series of continuum brightness temperature maps. Please note that data sets for (quasi-static) mosaics  can be retrieved directly from the ASA. For the data sets on SALSA, fundamental information such as date, band, helioprojective coordinates at the start of the time series, and identified co-observations are listed. In the first release, mostly co-observations with  IRIS are considered as it is a highly complementary observatory, providing sampling of the transition region and chromosphere in high spectral resolution. Attempts by multiple observatories at coordinating with ALMA were made but not necessarily guaranteed or successful due to the common challenges of observations such as accurate pointing and timing. This was the case also for a few observations coordinated with IRIS   that we found to not have overlapping target regions on the Sun.  For example, for D22 the disk centre was observed by ALMA which was not the coordinated target observed by IRIS. For D11, D16, D18 and D19 the correct targets were observed by IRIS but not at the same time as ALMA. A direct link to the specific IRIS observations page is provided on SALSA's web-interface, which typically includes  information on further observatories, including SDO and Hinode. The SALSA web-interface also provides the corresponding full-disk TP map of the Sun on which 
%, the table presents 
the location of the interferometrically observed region is marked. 
%drawn against a TP map showing the full Solar disk. 
These are the TP maps used for calibration (multiple TP observations may exist per set). Such TPs are available from ASA, but are reproduced in SALSA for convenience and record completeness. As an exception, being an early set, D01's TP is only available in raw format at ASA but was processed in the same way as the remaining TPs for the production of D01 (described in Sect.~\ref{sec:process}), and is available as a final product at SALSA.%, for the production of D01. %Except for D01, such TPs are also available at ASA, but are here made available for convenience and record completeness (including the TP used for D01).    

Some sets from SALSA have been reduced by other researchers independently from this database and were used in diverse publications. The reductions are not always comparable. A main point of difference is that we only release  sets at high cadence and other publications might consider averaged properties. %Despite differences in processing, the average temperature values found by other researchers agree within less than a hundred Kelvin with the corresponding values in this release.
The last column of Table~\ref{tab:sets} identifies all publications known to have used the same raw data (but with different processing) and publications for which sets from this database were used. The latter are marked in bold.  

%The sets from project ID 2016.1.00572.S, 2017.1.00653.S, 2017.1.017633.S  were used for \cite{2020A&A...640A..57A,2021ApJ...906...83C}. \textbf{The sets from ...}  % Our average temperature differs by 1xx K. 

%  Although these sets are made publicly available by ASA, contact with the PIs of the project is advised for best scientific results. 

\begin{sidewaystable*}
%\begin{landscape}
%\begin{table}
\centering
\begin{tabular}{llllclllcccll}
Data & Date & Project ID & Band / $\lambda$ & Cad.  & Obs. Time (UTC) & $\mu$ & T mean {[}K{]} & Mean resolution\footnote{Time average of the minor and major axes of the clean beam.} & Co- & Coordinates \footnote{necessarily approximate, see text} & Related \\
 & & & & {[}sec{]} & & & & bmin / bmaj [arcsec] & Observations & & publications \footnote{Publications where the SALSA set was used and publications where the data were used but processed independently from this release} \\
D01 & 2016-12-22 & 2016.1.00423.S & 3 / 3.0 mm & 2 & 14:19:31-15:07:07 & 0.99 & 7387 ± 519  &   1.37 / 2.10 & SDO              & 0,0 & {\bf 1}, {\bf 2}, 3, {\bf 4}                  \\
D02 & 2017-04-22 & 2016.1.00050.S & 3 / 3.0 mm & 2 & 17:20:13-17:42:37 & 0.92 & 9317 ± 1229  &  1.69 / 2.21 & IRIS,SDO             & -246,267 & 3, {\bf 4}, 5, 6, 7, {\bf 15}                     \\
D03 & 2017-04-23 & 2016.1.01129.S & 3 / 3.0 mm & 2 & 17:19:19-18:52:54 & 0.96 & 7161 ± 1564   & 1.92 / 2.30 & IRIS,SDO              & -54,251 & 3, {\bf 4}, 8                     \\
D04 & 2017-04-27 & 2016.1.01532.S & 3 / 3.0 mm & 2 & 14:19:52-15:31:17 & 0.78 & 7974 ± 1145   & 1.74 / 2.23 & IRIS,SDO              & 520,272 & 3, {\bf 4}                     \\
D05 & 2017-04-27 & 2016.1.00202.S & 3 / 3.0 mm & 2 & 16:00:30-16:43:56 & 0.96 & 7287 ± 1297  &  1.77 / 1.88 & IRIS,SDO              & 172,-207 & 3, {\bf 4}, 9, 10                   \\
D06 & 2018-04-12 & 2017.1.00653.S & 3 / 3.0 mm & 1 & 15:52:28-16:24:41 & 0.90 & 7586 ± 661   &  1.77 / 2.55 & IRIS,SDO             & -128,400 &{\bf 4}, 11                     \\
D07 & 2017-04-18 & 2016.1.01129.S & 6 / 1.3 mm & 2 & 14:22:01-15:09:15 & 0.76 & 7167 ± 1158   &  0.75 / 2.03  & IRIS,SDO              & -573,230 & 3, {\bf 4}, 8                    \\
D08 & 2017-04-22 & 2016.1.00050.S & 6 / 1.3 mm & 2 & 15:59:17-16:43:26 & 0.92 & 7496 ± 1014   &  0.68 / 0.85  & IRIS,SDO             & -261,266 & 3, {\bf 4}, 5, 6, 7                    \\
D09 & 2018-04-12 & 2017.1.00653.S & 6 / 1.3 mm & 1 & 13:58:58-14:32:27 & 0.88 & 5700 ± 333   &  0.80 / 2.22  & IRIS,SDO             & -175,-415 &{\bf 4}, 11                      \\
D10 & 2018-08-23 & 2017.1.01672.S & 6 / 1.3 mm & 1 & 16:24:27-17:18:05 & 0.97 & 6104 ± 497   &  1.69 / 2.21  & IRIS,SDO              & 68,-211 & {\bf 4}                  \\
D11 & 2017-03-16 & 2016.1.00572.S & 3 / 3.0 mm & 1 & 15:22:33-15:32:37 & 0.00 & 7263* ± 148  &  2.55 / 4.43  & SDO              & -679,-679 & 3, 11, 12, 13, 14  \\
D12 & 2017-03-19 & 2016.1.00030.S & 3 / 3.0 mm & 2 & 18:16:02-19:10:13 & 0.86 & 7364 ± 302  &   2.52 / 5.06  & IRIS,SDO             & -513,-64 & -- \\
D15 & 2018-08-23 & 2017.1.01672.S & 6 / 1.3 mm & 1 & 17:37:00-18:15:57 & 0.87 & 5814 ± 267  &   0.82 / 1.31  &  SDO                & 79,-238 &  {\bf 4}\\
D16 & 2017-03-16 & 2016.1.00572.S & 3 / 3.0 mm & 2 & 16:58:00-17:06:40 & 0.59 & 7768 ± 186  &   2.70 / 4.32  & SDO              & -517,-585 & 3, 11, 12, 13, 14                      \\
D17 & 2017-03-16 & 2016.1.00572.S & 3 / 3.0 mm & 2 & 18:40:51-18:50:56 & 0.84 & 7510 ± 241  &   2.54 / 5.17  & IRIS,SDO              & -321,-404 & 3, 11, 12, 13, 14                     \\
D18 & 2017-03-16 & 2016.1.00572.S & 3 / 3.0 mm & 2 & 17:59:04-18:09:08 & 0.72 & 7296 ± 251  &   2.52 / 4.85  & SDO              & -468,-484 & 3, 11, 12, 13, 14                    \\
D19 & 2017-03-16 & 2016.1.00572.S & 3 / 3.0 mm & 2 & 19:23:00-19:33:05 & 0.91 & 7416 ± 234  &   2.46 / 5.77  & SDO              & -261,-295 & 3, 11, 12, 13, 14                      \\
D20 & 2017-03-16 & 2016.1.00572.S & 3 / 3.0 mm & 2 & 16:14:48-16:24:53 & 0.13 & 4475 ± 256  &   2.49 / 4.56  & SDO              & -686,-666 & 3, 11, 12, 13, 14                      \\
D21 & 2018-12-20 & 2018.1.01763.S & 6 / 1.3 mm & 1 & 13:19:19-14:07:32 & 0.38 & 6181 ± 208  &   0.60 / 1.05  & IRIS,SDO             & 888,203 & 11                     \\
D22 & 2018-12-22 & 2018.1.01879.S & 6 / 1.3 mm & 1 & 15:09:14-15:14:57 & 1.00 & 6310 ± 185  &   0.73 / 1.99  & SDO \footnote{DC not targeted by IRIS} & 1, 0 & --                     \\
D23 & 2017-04-23 & 2016.1.01129.S & 6 / 1.3 mm & 2 & 14:23:55-15:11:06 & 0.41 & 6643 ± 396  &   0.71 / 1.72  & SDO             & -860,-129 & 3, {\bf 4}, 8                      \\
D24 & 2017-03-19 & 2016.1.00030.S & 3 / 3.0 mm & 2 & 15:32:32-16:26:52 & 0.84 & 7461 ± 421  &   2.65 / 4.39  & IRIS,SDO             & -535,-66 & --  \\
D25 & 2017-03-19 & 2016.1.00030.S & 3 / 3.0 mm & 2 & 16:52:52-17:47:11 & 0.85 & 7387 ± 415  &   2.48 / 4.31  & IRIS,SDO             & -484,-46 & --  \\
D26 & 2018-04-12 & 2017.1.00653.S & 3 / 3.0 mm & 1 & 16:43:52-17:16:06 & 0.90 & 7553 ± 360  &   1.84 / 2.82  & IRIS,SDO             & -131,-400 &{\bf 4}, 11, 16                      \\
D27 & 2018-04-12 & 2017.1.00653.S & 6 / 1.3 mm & 1 & 14:51:20-15:25:01 & 0.90 & 5899 ± 157 &    0.89 / 2.01  & IRIS,SDO             & 145,-400 &{\bf 4}, 11, 16                     \\
D28 & 2017-03-28 & 2016.1.00788.S & 6 / 1.3 mm & 1 & 15:09:20-16:12:12 & 0.91 & 6157 ± 162  &   1.05 / 1.89  & IRIS,SDO             & -181,347 & 3                    
\end{tabular}
\caption{Table listing all datasets present and downloadable in the initial release of SALSA. The date and time of the observations, the associated project ID, receiver band, wavelength, cadence, coordinates and $\mu$ angle, mean temperature and respective standard deviation, average resolution given as the time-averaged minor and major axes of the synthesised beam, identified co-observations by IRIS and publications where the set was used as well as publications where the same data were used but processed independently are listed. Publications that used the sets as processed in this database as listed in bold. Note that coordinates are necessarily approximate and the accuracy depends on the set and visually identifiable features (see text). {\bf Refs: }1: \citet{2020A&A...635A..71W}; 2: \citet{2020A&A...644A.152E}; 3: \citet{2021ApJ...910...77M}; 4: \citet{2021RSPTA.37900174J}; 5: \citet{2020A&A...634A..56D}; 6: \citet{2021ApJ...906...82C}; 7: \citet{2021ApJ...906...83C}; 8:  \citet{2019ApJ...881...99M}; 9: \citet{2019ApJ...877L..26L}; 10: \citet{2020ApJ...891L...8M}; 11: \citet{2020A&A...640A..57A}; 12: \citet{2018A&A...619L...6N}; 13: \citet{2020A&A...634A..86P}; 14: \citet{2020A&A...638A..62N}; 15: \citet{2021RSPTA.37900184G}; 16: \citet{2021A&A...652A..92N}
}
\label{tab:sets}
%\end{table}
%\end{landscape}
 \end{sidewaystable*}
%

%-------------------------------------------------------------------
\section{Obtaining and using SALSA data}
\label{sec:salsa-salat}

\subsection{SALSA}

The data files in FITS format can be downloaded from SALSA\footnote{Please send requests for technical support or scientific collaboration to the \email{solaralma@astro.uio.no}\label{fn:contact}.} 
by accessing \url{http://sdc.uio.no/salsa/} and clicking ``download'' in the appropriate column for each of the desired sets. Sets can be selected using their listed properties, manually or by using the search bar on the top-right of the table, by the quality and visible structures upon examination of the playable embedded movie (press ``movie'' button), or by the location on the solar disk and respective context as identifiable by the square marker overlaid on the image of the TP single-dish scan shown in the last column. The latter corresponds to the TP map  used for total power calibration as described in Sect.~\ref{sec:process}. The data arrays are ready to use in a form analogous to any other solar data and follow the format described in Sect.~\ref{sec:data}, with the headers and extensions providing additional advanced information following the standard described in Appendix~(\ref{sec:datahead} and \ref{sec:extension_data}).       

\subsection{SALAT}
\label{sec:salat}

The Solar ALMA Library of Auxiliary Tools (SALAT$^{\,\ref{fn:contact}}$) enables easy loading and initial visualisation/exploration of SALSA data products in both IDL and Python. For a complete description of SALAT, installation guide and examples, please visit \url{https://solaralma.github.io/SALAT/}. SALAT includes routines (IDL) and functions (Python) for reading the FITS files and extracting useful information such as arrays with the observing times or the synthesised beam shape for each ALMA image frame. It can be used to compute basic statistics for the time-series or individual frames as well as for the corresponding ALMA synthesised beam, important for advanced analysis. Functions to plot ALMA data, including the beam shape, and the possibility of saving the images are included. SALAT can be used as well for identifying the best frames in an observation or to get a general idea of the timeline during the observation period, including any time gaps due to filtered out frames or concatenated scan periods. Useful for the production of synthetic observables and comparison with co-observations, SALAT allows the usage of the ALMA beam of a particular observation to convolve it with other types of data, such as complementary datasets and simulations. Finally, SALAT can be used to create a new FITS file with the appropriate format (e.g. reduced dimensions), which can be inspected with other external viewing/analysis tools as SAOImageDS9 \citep{2003ASPC..295..489J}, CARTA \citep{angus_comrie_2021_4905459}, CRISPEX \citep{2012ApJ...750...22V,2018arXiv180403030L}, etc.

\subsection{Acknowledgements and optional collaboration}

We kindly ask to add the following acknowledgement to any publication that uses data downloaded from SALSA: 

\textit{``This paper makes use of the following   ALMA   data:} [ALMA-PROJECT-ID]~\textit{. ALMA   is   a   partnership   of   ESO   (representing  its  member  states),  NSF  (USA)  and  NINS  (Japan),together with NRC(Canada), MOST and ASIAA (Taiwan), and KASI  (Republic  of  Korea),  in  co-operation  with  the  Republic of  Chile.  The  Joint  ALMA  Observatory  is  operated  by  ESO, AUI/NRAO and NAOJ. 
SALSA, SALAT and SoAP are produced and maintained by the SolarALMA project, which has received  funding  from  the  European  Research  Council  (ERC) under the European Union’s Horizon 2020 research and innovation  programme  (grant  agreement  No.  682462),  and  by  the Research Council of Norway through its Centres of Excellence scheme, project number 262622.''} 

The ALMA-PROJECT-ID can be found in the FITS header keyword PROJID of the SALSA data file, e.g.  ADS/JAO.ALMA\#2016.1.00423.S.  

Optionally, we are open for collaboration on science projects in the form of technical support and/or help with scientific analysis and co-authorship of resulting publications. Please refer to the SALSA webpage\footnote{\url{http://sdc.uio.no/salsa/}} for more information or contact us directly$^{\,\ref{fn:contact}}$.

%-------------------------------------------------------------------
\section{Summary and Outlook}
\label{sec:outlook}

%Here we present the first release of SALSA - a database of science-ready data sets for ALMA observations of the Sun. 
Here we present the first release of SALSA - a database of 26 science-ready data sets for ALMA observations of the Sun. We also present the first release of SALAT, an IDL/Python package for loading and initial visualisation/exploration of SALSA data products.
%The details for the datasets that are included in this first version of the SALSA are provided. 
We plan to extend SALSA with additional datasets in the future. Depending on the further development of solar observing modes with ALMA, new data sets for new capabilities will be added. 
%with specifics extending outside what is described here might be added, in particular as the offered capabilities for solar ALMA observations are developed. 
That includes for instance data measured in other receiver bands (e.g., bands~5 or 7), different frequency setup (e.g., sub-band sampling), other antenna array configurations but also different image reconstruction techniques.

%-------------------------------------------------------------------
\section*{Acknowledgments}
This work is supported by the SolarALMA project, which has received funding from the European Research Council (ERC) under the European Union’s Horizon 2020 research and innovation programme (grant agreement No. 682462), and by the Research Council of Norway through its Centres of Excellence scheme, project number 262622.
This paper makes use of the following ALMA data: 
ADS/JAO.ALMA\#2016.1.00030.S, 
ADS/JAO.ALMA\#2016.1.00050.S, 
ADS/JAO.ALMA\#2016.1.00202.S, 
ADS/JAO.ALMA\#2016.1.00423.S, 
ADS/JAO.ALMA\#2016.1.00572.S, 
ADS/JAO.ALMA\#2016.1.01129.S, 
ADS/JAO.ALMA\#2016.1.01532.S, 
ADS/JAO.ALMA\#2017.1.00653.S, 
ADS/JAO.ALMA\#2017.1.01672.S, 
ADS/JAO.ALMA\#2018.1.01879.S, and 
ADS/JAO.ALMA\#2018.1.01763.S. 
ALMA is a partnership of ESO (representing its member states), NSF (USA) and NINS (Japan), together with NRC(Canada), MOST and ASIAA (Taiwan), and KASI (Republic of Korea), in co-operation with the Republic of Chile. 
The Joint ALMA Observatory is operated by ESO, AUI/NRAO and NAOJ. We are grateful to the many colleagues who contributed to developing the solar observing modes for ALMA, including the international ALMA solar development group and the participants of the First International Workshop on Solar Imaging with ALMA (ALMA-SOL-IMG1), and for support from the ALMA Regional Centres. We are indebted to the ALMA Science Archive team for their excellent product and to Felix Stoehr in particular for comments on the FITS header standard. We thank Pit S\"{u}tterlin for his excellent code contributions throughout the years which we continuously come back to when building new infrastructure such as image alignment and scaling optimisation routines. Rob Rutten's public SDO alignment routines were used for the figure examples. We thank Terje Fredvik for valuable comments on the final header products and Stian Aannerud for the final testing of SALSA and SALAT.

%---------------------------------------------------------------
%-------------------------------------------------------------------
%-------------------------------------------------------------------

\bibliographystyle{aa}
\bibliography{bib.bib}
%-------------------------------------------------------------------
%-----------------------
\begin{appendix}
\onecolumn

\section{Data file headers} 
\label{sec:datahead}

The header keywords in the SALSA FITS files seek to ensure reproducibility of the data, facilitate easy search across the archive for data fulfilling specific criteria, and  availability of critical metadata information in a standardised fashion that respects all previous conventions. We specifically guarantee that the SALSA data format conforms to the FITS standard 3.0 and 4.0, the SOLARNET recommendations \citep[under revision:][]{2020arXiv201112139H}, representation of World Coordinates in FITS \citep{2002A&A...395.1061G}, World Coordinate System for Solar Observations for Solar observations \citep[following][]{2006A&A...449..791T}, and \citet{2015A&A...574A..36R} for time coordinate representation in FITS files. The most recent evolving ALMA FITS data-product standard (Felix Stoehr, private communication) and ALMA memo 611 were also followed with slight departures from the memo where a collision was present. ALMA memo 613 was also used to inform decisions where two options were possible. Adaptation of existing keywords was necessary but kept to a minimum and never breaking any standard. Three headers exist, one per HDU, with relevant keywords per extension.

The keywords are listed, per extension, in Tables~\ref{tab:SALSAhead} and \ref{tab:cube_ext} with an example and a short description. The helioprojective coordinates keywords are summarised in Table~\ref{tab:wcs}. In the description of each keyword we include the list of standards that constrain or originated its usage. These are abbreviated in the following way: F3.0 for NASA FITS Standard 3.0 (sometimes preferred in CASA documentation), F4.0 for NASA FITS Standard 4.0, T2006 for Thompson (2006), SNET for the SOLARNET recommendations document, S1.9 for the ALMA FITS data-product standard version 1.9, and A611 or A613 for ALMA memos 611 and 613, respectively. 

World Coordinate System keywords are used according to \citet{2002A&A...395.1061G} and \citet{2006A&A...449..791T} to describe the data coordinates in two separate coordinate systems. The primary coordinate system specifies RA/DEC in a SIN projection, thus CTYPE1='RA---SIN' and CTYPE2='DEC--SIN'. The secondary coordinate system specifies helioprojective longitude and latitude in a gnomonic projection, thus CTYPE1A = 'HPLN-TAN' and CTYPE2A = 'HPLT-TAN' (note suffix A in keyword names for the secondary system).

The existing CASA keywords are passed through at this stage unless they are set to be changed as per the most recent information regarding the ALMA FITS data-product standard. An example of this is the keyword ORIGIN, set in most archival data to be the CASA version where it should instead display the originating institution (in agreement with FS3.0 and FS4.0). In such cases, the header in SALSA data files is corrected.

\noindent
\begin{table}[h!]
    \centering
%\begin{tabular}{|l|c|l|}
\begin{tabular}{ | c | m{6cm} | m{8cm} | }
\hline
\multicolumn{3}{|c|}{Generic FITS keywords common for all SALSA data products}\\
\hline
Keyword&Example Value&Description\\
\hline
\texttt{PROPCODE}& '2018.1.01879.S' & Project code to which the data belongs. String. Set by S.19. \\
\hline
\texttt{PROJECT}& 'SolarALMA' & Project. String. Set by SNET.  \\
\hline
\texttt{INSTRUME}& 'BAND3' & Band. String. Reserved by FS3.0. FS4.0. Set by S1.9. \\
\hline
\texttt{FILENAME}& 'solaralma.....image.fits' & Filename. String. Set by SNET.   \\
\hline
\texttt{DATAMIN}& 5800 & Minimum value of the data. Float. Reserved and set by: F3.0, F4.0, SNET.      \\
\hline
\texttt{DATAMAX}& 6100 & Maximum value of the data. Float. Reserved and set by: F3.0, F4.0, SNET.    \\
\hline
\texttt{CASAVER}& 'CASA 3.4.0(release r19988)' & String. Set by S1.9 and A613. \\
\hline
\texttt{ORIGIN}& 'JAO-UIO' & UIO appended. String. Reserved by F3.0. F4.0. Set by S1.9.\\
\hline
\texttt{SOLARNET}& '0.5' & 0.5 or 1 for partially compliant or fully compliant. String. Set by SNET \\
\hline
\texttt{PWV}& 0.7 & Unit: mm. Average precipitable water vapour. Float. A quality metric.  \\
\hline
\texttt{SPATRES}& 0.5 & Geometric average of the min and the max beam axes (a sort of "Spatial resolution").  Float. Arcsec. Set by A613 and S1.9. \\ 
\hline
\texttt{BNDCTR} & 2.315424966698E+11 & Center frequency of data in the FITS array. Hz. Set by S1.9 and SNET. \\
\hline
\texttt{WAVEBAND} & 'ALMA Band 3' & Human readable band description. SNET. \\
\hline
\texttt{OBS\_HDU}& 1 & Contains Observational data. Integer. Set by SNET. \\
\hline
\texttt{DATE-BEG}& '2014-12-11T19:09:13' & Start of solar observations. String. Set by FS3.0 ,FS4.0, SNET. \\
\hline
\texttt{DATATAG} & \footnotesize{'This paper makes use of the following ALMA data: ADS/JAO.ALMA\# [Project
code]. ALMA is a partnership of ESO (representing its member states), NSF (USA) and NINS
(Japan), together with NRC (Canada) and NSC
and ASIAA (Taiwan), in cooperation with the Republic of Chile. The Joint ALMA Observatory is
operated by ESO, AUI/NRAO and NAOJ. It also makes use of the Solar ALMA Science Archive developed at UiO and based on SoAP reduction. Please cite ...'} &  Set by S1.9. Implemented in SALSA as an OGIP 1.0 compliant LONGSTRN.  \\
\hline
\end{tabular}
    \caption{A list of keywords used in all SALSA data products and respective description with examples. Part I.  Continued in Table~\ref{tab:SoAPhead}. }
    \label{tab:SALSAhead}
\end{table}
\vspace*{\baselineskip}

\noindent
\begin{table}[h!]
    \centering
\begin{tabular}{ | c | c | m{8cm} | }
\hline
\multicolumn{3}{|c|}{Primary HDU FITS keywords}\\
\hline
\hline
Keyword&Default Value&Description\\
\hline
\texttt{CREATOR}& 'SoAP' & Name of the last software that created the data contents. String. Set by SNET.\\
\hline
\texttt{VERS\_SW}&'0.12'& Version of the CREATOR software. String. Set by SNET. \\
\hline
\texttt{HASH\_SW} & '424d417e04...' & Hash of CREATOR software (robust versioning). Can be GIT hub commit hash. Example is the hash for the latest version marked as v0.14. String. Making its way into SNET. \\ 
\hline
\texttt{CO\_OBS}& 'IRIS 3640009423,SST,NST' & Comma separate list of telescopes with identified co-observations and strings of observation IDs if available. String. \\
\hline
\texttt{CHNRMS}& 0.324E-03 & Channel RMS. Kelvin. Float. Set by S1.9 and A613 (but not in Jy/beam).  \\
\hline
\texttt{FOV}& 1.54 & Total area of the field of view of the image. Square degrees. Float. Set by S1.9. \\ % does not make much sense to have this and EFFDIAM but we follow S1.9
\hline
\texttt{EFFDIAM} & 0.70 &  Effective diameter of the field of view. Degrees. Float. Set by S1.9\\
\hline
\texttt{RES\_MEAN}& 0.1 & Mean of residuals after CLEAN. Float. Basic quality measure.  \\
\hline
\texttt{RES\_STDV}& 0.1 & Standard deviation of the residuals. Float. Basic quality measure. \\
\hline
\texttt{VAR\_KEYS} & 'TIMEVAR; BMAJ, BMIN, BPA, TIME\_TAG' &  Reference to extension containing time-variable values. String. Set by SNET.\\
\hline
\texttt{PRSTEP1} & 'CLEAN' &  Processing steptype. Generic form is PRSTEPn where n stands for number of processing step. String. Set by SNET.\\
\hline
\texttt{PRPROC1} & 'TCLEAN' &   Name of procedure performing PRSTEP1. String. Set by SNET.\\
\hline
\texttt{PRPVER1} &    0.8  & Version of procedure PRPROC1. String. Set by SNET.\\
\hline
\texttt{PRREF1A} & 'UIO-ALMA (solaralma@uio.no)' & Group doing manual adjustments. String. Set by SNET.\\
\hline
\texttt{PRPARA1} & '[PRPARA]' & Name of extension containing list of parameters/options for PRPROC1. String. Set by SNET.\\
\hline
\texttt{PRLIB1}  & 'CASA  ' & Software library containing TCLEAN. String. Set by SNET. \\
\hline
\texttt{PRVER1}  &    5.6 & Library version/MJD of last update. String. Set by SNET.\\
\hline
\texttt{PRSTEP2} & 'TPCALIB' &  PRocessing steptype. Generic form is PRSTEPn where n stands for number of processing step. Set by SNET.\\
\hline
\texttt{PRPROC2} & 'SOAP\_FEATHER' &   Name of procedure performing PRSTEP1. String. Set by SNET.\\
\hline
\texttt{PRPVER2} &    1.0  & Version of procedure PRPROC2. String. Set by SNET.\\
\hline
\texttt{PRPARA2} & '[PRPARA]' & Name of extension containing list of parameters/options for PRPROC2. String. Set by SNET.\\
\hline
\end{tabular}
    \caption{A list of keywords used in all SALSA data products and respective description with examples. Continued from \ref{tab:SALSAhead}}
    \label{tab:SoAPhead}
\end{table}
\vspace*{\baselineskip}

\noindent
\begin{table}[h!]
    \centering
\begin{tabular}{ | c | c | m{8cm} | }
\hline
\multicolumn{3}{|c|}{World Coordinate System - Helioprojective keywords }\\
\hline
Keyword&Default Value&Description\\
\hline
\texttt{WCSNAMEA}&'Helioprojective-cartesian'& Name of coordinate set. String. Set by T2006 and A611. \\
\hline
\texttt{CTYPE1A}&'HPLN-TAN'& Axis labels. Set by T2006 and A611. \\
\hline
\texttt{CTYPE2A}&'HPLT-TAN'& Axis labels. Set by T2006 and A611.  \\
\hline
\texttt{CRPIX1A}&180.5&Reference pixel: center of image in pixels. Set by T2006 . \\
\hline
\texttt{CRPIX2A}&180.5& Reference pixel: center of image in pixels. Set by T2006. \\
\hline
\texttt{CUNIT1A}&'arcsec'& Angles in arcsec. Set by T2006 and A611. Arcsec is allowed by WCS.   \\
\hline
\texttt{CUNIT2A}&'arcsec'& Angles in arcsec. Set by T2006 and A611. Arcsec is allowed by WCS.\\
\hline
\texttt{CDELT1A}& 0.55 & Plate scale (i.e. pixel size). Set by T2006. Here in arcsec as in A611.  \\
\hline
\texttt{CDELT2A}& 0.55 & Plate scale (i.e. pizel size). Set by T2006. Here in arcsec as in A611.    \\
\hline
\texttt{CRVAL1A}& 0.0 & Helioprojective x (solar x) coordinate of the reference pixel in arcsec. Set by T2006. Here in arcsec as in A611.  \\
\hline
\texttt{CRVAL2A}& 0.0 & Helioprojective y (solar y) coordinate of the reference pixel in arcsec. Set by T2006. Here in arcsec as in A611.   \\
\hline
\texttt{CTYPE5}& 'TIME'  & This is in relation to REF\_TIME, per frame. Time found in TIME\_TAG under extension TIMEVAR.  \\
\hline
\texttt{CUNIT5}& 's'  & Unit for fifth axis. Seconds in relation to REF\_TIME, per frame. Time found in TIME\_TAG under extension TIMEVAR. Normally preceding midnight. \\
\hline
\texttt{DSUN\_OBS}& 147296035413.001 & Distance between observer and Sun in meters. Set by T2006 and A611 with A611 precision.  \\
\hline
\texttt{RSUN\_REF}&696000000.0& Radius of the Sun in meters. Set by T2006 and A611 with A611 precision. \\
\hline
\texttt{RSUN\_OBS}& 6.96e8 & Observed radius of the Sun in arcsec. Set by T2006 and A611 with tabled A611 precision.  \\
\hline
\texttt{HGLN\_OBS}&0 &Heliographic longitude of the observer, in degrees. Set by T2006 and A611. \\
\hline
\texttt{HGLT\_OBS}& -0.47 & Stonyhurst heliographic latitude of the observer, in degrees. Set by T2006 and A611. \\
\hline
\texttt{SOLAR\_P} & 11.6872 & Degrees. "rotate image by -solar P angle to get solar north up". Set by A611. \\ % SOLAR\_P0 for SNET.
\hline
\texttt{SOLAR\_P0}& 11.6872 & Degrees. "rotate image by -solar P angle to get solar north up". Set by SNET. \\
\hline
\texttt{REF\_TIME}& 2014-12-11T19:09:13  & Date and time. Time used for coordinate calculations. Set by A611.  \\
\hline
\end{tabular}
    \caption{World coordinate system keywords for solar imaging data used in SALSA. Generic keywords.}
    \label{tab:wcs}
\end{table}
\vspace*{\baselineskip}

\section{Extension data}
\label{sec:extension_data}

We use extensions for technical details regarding data reduction and time-varying information. The extension ``TIMEVAR'' contains time varying information stored in a standard binary table format. It follows the variable-keyword mechanism described in Appendix I-d of the SOLARNET recommendations. It can be selected programmatically by name: \texttt{EXTNAME = 'TIMEVAR'}, and is referenced in the main header through the VAR\_KEYS keyword. Its parameters are listed in Table~\ref{tab:cube_ext}. These are the beam shape parameters BMAJ, BMIN, and BPA as a function of time and the time stamps (TIME\_TAG) of each frame of the primary HDU. The TIME\_TAG values are expressed in seconds from DATEREF following \citet{2015A&A...574A..36R}. 

% This extension is  binary in compliance with Appendix I of SNET and \cite{1995A&AS..113..159C}
\begin{table}[h!]
 \centering
\begin{tabular}{ | m{2cm} | c | m{8cm} | }
\hline
\multicolumn{3}{|c|}{Header of the extension containing time dependent variables. }\\
\hline
Keyword&Default Value&Description\\
\hline
\texttt{XTENSION} & 'BINTABLE' &  Type of extension. String. Set by F4.0. \\
\hline
\texttt{TFIELDS} & 4 &  Total fields. Integer. Set by F4.0. \\
\hline
\texttt{EXTNAME} & 'TIMEVAR' & Name of the extension. String. Primary HDU refers to this name using VAR\_KEYS \\
\hline
\texttt{DATEREF} & '2014-12-11T00:00:00' & Time in UTC. Reference time that sets the zero to all other UTC times such as the extension time stamps. String. Set by SNET and F4.0. \\
\hline
\texttt{SOLARNET}& 0.5 & SOLARNET compliance: 0.5 or 1 for partially compliant or fully compliant. Float. Set by SNET.\\
\hline
\texttt{TTYPE1} & 'BMAJ' &  First column. Beam shape major axis. Variable-keyword. String. Degrees. Set by F4.0. \\
\hline
\texttt{TTYPE2} & 'BMIN' &  Second column. Beam shape minor axis. Variable-keyword. String. Degrees. Set by F4.0. \\
\hline
\texttt{TTYPE3} & 'BPA' & Third column. Beam average. Variable-keyword. String. Degrees. Set by F4.0. \\
\hline
\texttt{TTYPE4} & 'TIME\_TAG' &  Fourth column. Tabulation of time. Variable-keyword. String. Set by F4.0. \\ %type time also works as controlled by main header
\hline
\texttt{TFORM1} & '1558E' &  Real*4 (float). String. Set by F4.0 and SNET. \\
\hline
\texttt{TFORM2} & '1558E' &  Real*4 (float). String. Set by F4.0 and SNET. \\
\hline
\texttt{TFORM3} & '1558E' &  Real*4 (float). String. Set by F4.0 and SNET. \\
\hline
\texttt{TFORM4} & '1558E' &  Real*4 (float). String. Set by F4.0 and SNET.  \\ 
\hline
\texttt{TDIM1} & '(1,1,1,1,1558)' &  Dimension of array. String. Set by F4.0 and SNET.  \\ 
\hline
\texttt{TDIM2} & '(1,1,1,1,1558)' &  Dimension of array. String. Set by F4.0 and SNET.  \\ 
\hline
\texttt{TDIM3} & '(1,1,1,1,1558)' &  Dimension of array. String. Set by F4.0 and SNET.  \\ 
\hline
\texttt{TDIM4} & '(1,1,1,1,1558)' &  Dimension of array. String. Set by F4.0 and SNET.  \\ 
\hline
\texttt{TUNIT1} & 'arcsec' &  Units for column 1. String. Set by F4.0 and SNET. \\
\hline
\texttt{TUNIT2} & 'arcsec' &  Units for column 2. String. Set by F4.0 and SNET. \\
\hline
\texttt{TUNIT3} & 'arcsec' &  Units for column 3. String. Set by F4.0 and SNET. \\
\hline
\texttt{TUNIT4} & 's' &  Units for column 4. String. Set by F4.0 and SNET.  \\
\hline
\end{tabular}
 \caption{Header of the extension with time dependent variables, extension ``TIMEVAR'', with example values. Note that the header keywords describe the variables in the binary table.}
    \label{tab:cube_ext}
\end{table}
\vspace*{\baselineskip}

The second extension, ``PRPARA'', contains parameters that might change with dataset for each processing step, stored as an ASCII table for readability. Code box~\ref{verb1} presents a print out example of such ASCII table from a specific file. The format is compatible with F3.0 and F4.0, and is included for the goal of reproducibility. The main header merely refers to that extension by its name in square brackets in the PRPARAnn keyword, as defined in the SOLARNET recommendations (described in Section 8.2 of the version currently under revision). The main header PRSTEPn keyword is used to specify the main methods used, for which the parameters in this ASCII table will be relevant. Table~\ref{Extension_header} lists an example of a header of an extension containing such processing parameters. Note these include a repetition of the PRxxxx keywords as also suggested by the SOLARNET recommendations. In the example printed out, the first 17 rows are spelled as in the CASA tclean routine call, as made by SoAP, and are recognisable in the general context of the CLEAN algorithm (the headers PRSTEP1). The remaining rows refer to other SoAP post-processing steps using native functions (the headers PRSTEP2). 

\begin{table}[!h]
%\begin{tabular}{ | c | m{3cm} | m{8cm} | }
 \centering
\begin{tabular}{ | c | m{4cm} | m{8cm} | }
\hline
\multicolumn{3}{|c|}{Extension data}\\
\hline
Keyword&Default Value&Description\\
\hline
%\texttt{Residuals}& Float two dimensional array for single images and three dimensional array for cubes & Residuals from tclean \\
%\hline
\texttt{EXTNAME}& \centering'PRPARA' & Name of extension. This keyword is recognized by multiple libraries.  String. Making its way to SNET.  \\
\hline
\texttt{PRSTEP1} &  \centering'CLEAN' &  Processing step type. Generic form is PRSTEPn where n stands for number of processing step. String. Set by SNET.\\
\hline
\texttt{PRPROC1} &  \centering'TCLEAN' &   Name of procedure performing PRSTEP1. String. Set by SNET.\\
\hline
\texttt{PRPVER1} &  \centering   '0.8'  & Version of procedure PRPROC1. String. Set by SNET.\\
\hline
\texttt{PRREF1A} &  \centering'UIO-ALMA (solaralma@astro.uio.no)' & Group doing manual adjustments. String. Set by SNET.\\
\hline
\texttt{PRPARA1} &  \centering'[PRPARA]' & Name of extension containing list of parameters/options for PRPROC1/PRPROC2. String. Set by SNET.\\
\hline
\texttt{PRLIB1}  &  \centering'CASA  ' & Software library containing TCLEAN. String. Set by SNET. \\
\hline
\texttt{PRVER1}  &  \centering   '5.6' & Library version/MJD of last update. String. Set by SNET.\\
\hline
\texttt{PRSTEP2} & \centering 'TPCALIB' &  Processing steptype. Generic form is PRSTEPn where n stands for number of processing step. String. Set by SNET.\\
\hline
\texttt{PRPROC2} & \centering 'SoAP+' &   Name of procedure performing PRSTEP2. String. Set by SNET.\\
\hline
\texttt{PRPVER2} & \centering    '1.0'  & Version of procedure PRPROC2. String. Set by SNET.\\
\hline
\texttt{PRREF2A} & \centering 'UIO-ALMA (solaralma@astro.uio.no)' & Group doing manual adjustments. String. Set by SNET.\\
\hline
\texttt{PRLIB2}  &  \centering'SoAP' & Software library containing SALAT+. String. Set by SNET. \\
\hline
\texttt{PRVER2}  & \centering   '1.00' & Library version/MJD of last update. String. Set by SNET.\\
\hline
\texttt{PRPARA2} & \centering'[PRPARA]' & Name of extension containing list of parameters/options for PRPROC1/PRPROC2. String. Set by SNET.\\
\hline
\end{tabular}\\
 \caption{List of SALSA-specific header keywords for the extension containing processing parameters. These are mostly repeated from table 4.3 following direct recommendation from SOLARNET and so that the main header does not need to be searched in order for the extension data to be useful.}
\label{Extension_header}
\end{table}
%\vspace*{\baselineskip}

\begin{table}[H]
    \renewcommand\tablename{Code}
    \begin{verbatim}
          cellsize [0.5 arcsec,0.5 arcsec]      Cellsize.String array input with units.
             Niter                  200000                        Number of iterations.
            robust                     0.5                               ROBUST. Float.
        cycleniter                    1000                         Cycleniter. Integer.
            pbmask                     0.3                               PBMASK. Float.
           pblimit                    True                        PBLIMIT. Always true.
           usemask                      pb                 String. Usually set to "pb".
              gain                   0.025                                 Gain. Float.
       deconvolver              multiscale                         DECONVOLVER. String.
       phasecenter                       0                   Always zero unless mosaic.
            scales                [0,6,18]                       Scales. Integer array.
         weighting                  briggs                           Weighting. String.
           uvtaper           [0.0,0.2,0.3]           Float array. Smoothing parameters. 
    smallscalebias                      ''      Smallscale bias. String, usually blank.
          normtype               flatnoise    Normalization type for residuals. String.
         threshold                    1 Jy           Stop threshold. String with units.
           gridder                mosaicft               Mosaic as two arrays (12m 7m).
         mosweight                    True                           Bool. Always true.
       TPCalMethod     White et al. (2018)                         Calib method for TP.
        TPCalValue                    7300                    Calib temperature for TP.
             REFTP  Xcc8b19_X4deb...manual        Identifier for calib TP. Abbreviated.
         CALIBTEMP                 7238.14                DC temperature level from TP.
         smoothfun                  BoxCar                    PTseries smooth function.
        windowsize                5 frames                      smooth w. size, frames.
       FiltMethods             RMS,Central                Methods for filtering frames.
             Nrefs                       4            Reference interval for alignment.
    \end{verbatim}
    \caption{An example print out from extension PRPARA (extension 2), containing processing parameters for different steps. Referred to from the headers.}
    \label{verb1}
\end{table}

\end{appendix}

\end{document}